\documentclass[fleqn,usenatbib]{mnras}

\usepackage[T1]{fontenc}
\usepackage{ae,aecompl}
\usepackage{siunitx}
\usepackage{pdflscape}
\usepackage{afterpage}
\usepackage{orcidlink}

\usepackage{graphicx}	
\usepackage{amsmath}	
\usepackage{amssymb}	

\usepackage{newtxtext,newtxmath}

\title[J1921: An Eccentric Binary]{Discovery of a Highly Eccentric, Chromospherically Active Binary: ASASSN-V J192114.84+624950.8}

\author[Z. Way et al.]{Zachary S. Way$^{1}$\thanks{E-mail: way.65@osu.edu},
T. Jayasinghe \orcidlink{0000-0002-6244-477X}$^{1,2}$\thanks{Ohio State Presidential Fellow},
C. S. Kochanek$^{1.2}$,
K. Z. Stanek$^{1,2}$,
\newauthor{
Patrick Vallely \orcidlink{0000-0001-5661-7155}$^{1}$\thanks{NSF Graduate Fellow},
Todd A. Thompson$^{1,2}$,
Thomas~W.-S.~Holoien \orcidlink{0000-0001-9206-3460}$^{3\thanks{NHFP Einstein Fellow}}$,} 
\newauthor{Benjamin J. Shappee \orcidlink{0000-0003-4631-1149}$^{4}$
}
\\
$^{1}$Department of Astronomy, The Ohio State University, 140 West 18th Avenue, Columbus, OH 43210, US\\
$^{2}$Center for Cosmology and AstroParticle Physics, The Ohio State University, 191 W. Woodruff Avenue, Columbus OH 43210, US \\
$^{3}$The Observatories of the Carnegie Institution for Science, 813 Santa Barbara St., Pasadena, CA 91101, USA\\
$^{4}$Institute for Astronomy, University of Hawai’i, 2680 Woodlawn Drive, Honolulu, HI 96822, USA
}

\date{Accepted XXX. Received YYY; in original form ZZZ}

\pubyear{2021}

\begin{document}
\label{firstpage}
\pagerange{\pageref{firstpage}--\pageref{lastpage}}
\maketitle

\begin{abstract}

As part of an All-Sky Automated Survey for SuperNovae (ASAS-SN) search for sources with large flux decrements, we discovered a transient where the quiescent, stellar source, ASASSN-V J192114.84+624950.8, rapidly decreased in flux by $\sim55\%$ ($\sim0.9$ mag) in the g-band. The \textit{TESS} light curve revealed that the source is a highly eccentric, eclipsing binary. Fits to the light curve using \textsc{phoebe} find the binary orbit to have $e=0.79$, $P_{\rm orb}=18.462~\text{days}$, and $i=88.6^{\circ}$ and the ratios of the stellar radii and temperatures to be $R_2/R_1 = 0.71$ and $T_{e,2}/T_{e,1} = 0.82$. Both stars are chromospherically active, allowing us to determine their rotational periods of $P_1=1.52$ days and $P_2=1.79$ days, respectively. A LBT/MODS spectrum shows that the primary is a late-G or early-K type dwarf. Fits to the SED show that the luminosities and temperatures of the two stars are $L_1 = 0.48~L_{\sun}$, $T_1= 5050~K$, $L_2 = 0.12~L_{\sun}$, and $T_{2} = 4190~K$. We conclude that ASASSN-V J192114.84+624950.8 consists of two chromospherically active, rotational variable stars in a highly elliptical eclipsing orbit.
\end{abstract}

\begin{keywords}
eclipsing binaries -- stars: chromospheres -- variable stars
\end{keywords}



\section{Introduction}

The All-Sky Automated Survey for SuperNovae \citep[ASAS-SN;][]{2014ApJ...788...48S, 2017PASP..129j4502K} is a survey designed to carry out unbiased searches for supernovae and other bright transient events. 
However, ASAS-SN simultaneously monitors the brightness of $\sim100$ million stars with $g\lesssim 18 ~\text{mag}$, allowing it to systematically identify and classify variable stars as well \citep{tharindu_variable_1, jayasinghe_2021}.  
Some types of variability are not automatically identified when searching for transients or normal variable stars.
In particular, stars that drop suddenly and infrequently in brightness can be missed.
Beginning in 2019, we began systematically searching for sources that dropped by $> 0.75$ mag. We have reported several unusual objects with such variability, including ASASSN-V J060000.76--310027.83 \citep{2019ATel13346....1W}, a likely R Coronae Borealis (RCB) variable \citep{2019ATel13159....1W}, and a star with deep, dimming episodes that does not appear to be an RCB candidate \citep{2019ATel13106....1W}.

One class of variable stars that is often missed by automated light curve searches is eclipsing binaries with long periods or very short eclipses. Two ASAS-SN examples of the former are an eclipsing giant star with a 750 day orbit \citep{2020ATel13745....1J} and a giant eclipsing binary with a 12 year orbit \citep{asassn_21co}. 
Here, we discuss an interesting example of the latter.  
ASASSN-V J192114.84+624950.8 (J1921 hereafter) is a binary consisting of two chromospherically active, rotational variable stars in a highly eccentric, 18 day orbit where the eclipses span only $\sim 2\%$ of the total phase.

J1921 was discovered as a $\Delta g=0.88$ mag dimming event on UT 2019-06-05.36, along with an earlier, un-flagged event on UT 2016-06-29.55 that also corresponds to the primary eclipse. 
Fortunately, J1921 also lies close to the Transiting Exoplanet Survey Satellite \citep[\textit{TESS};][]{tess_instrument_paper} continuous viewing zone (CVZ), providing a densely sampled light curve that fully samples the primary and secondary eclipses.

\begin{figure*}
	\includegraphics[width=\textwidth]{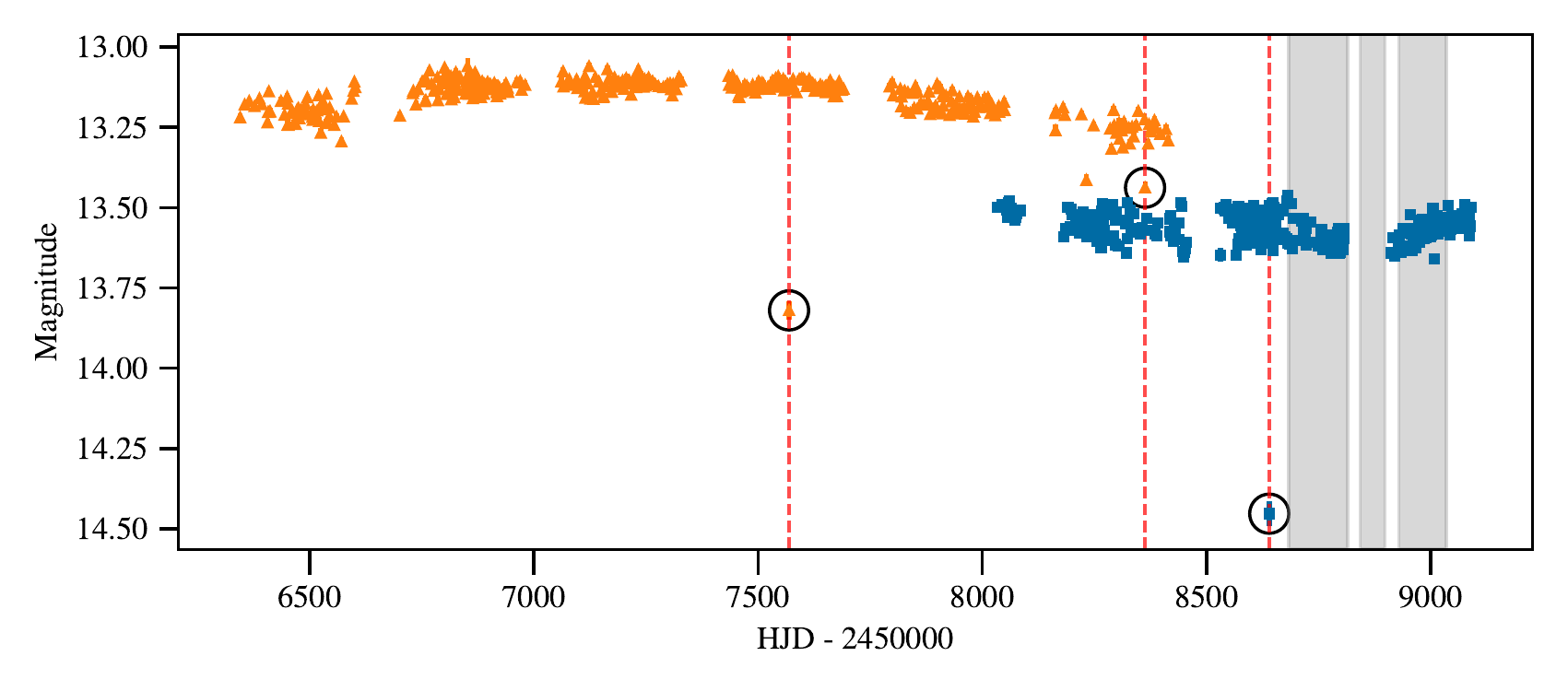}
    \caption{The ASAS-SN g-band (blue) and V-band (orange) light curves. The circled points represent the epochs that this star was dimmer than usual and they coincide with predicted eclipse times. The most recent example triggered our investigation. The primary eclipse times that coincided with ASAS-SN epochs are shown as red dashed lines. The gray stripes are the epochs spanned by \textit{TESS} data.}
    \label{fig:ASAS_lc}
\end{figure*}

\begin{figure*}
    \centering
    \includegraphics[width=0.95\textwidth]{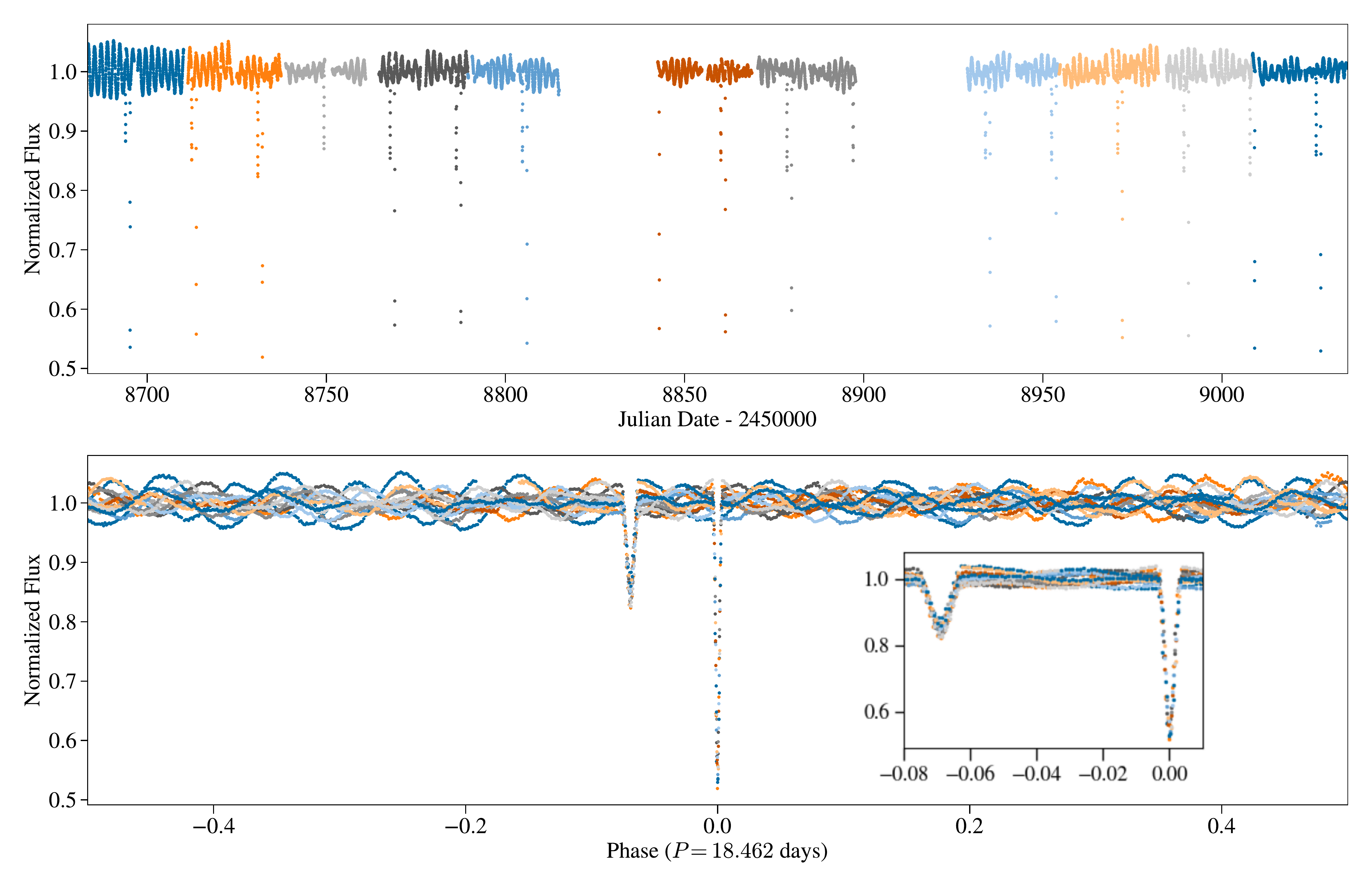}
    \caption{\textit{Top Panel}: The normalized \textit{TESS} light curve. \textit{Bottom Panel}: The light curve folded over the orbital phase with a window showing the eclipses in more detail.}
    \label{fig:Tess_lc}
\end{figure*}

We also detect two additional periodic signals in the \textit{TESS} light curve at $P\sim1.79$ and $P\sim1.52$ days. We attribute these to the rotation of the primary and secondary stars.
Rapidly rotating main sequence stars are generally assumed to be young \citep{kraft_rapid_rotation}. While rapid rotation is not a perfect metric for age, especially in binaries \cite[see][]{simonian_rapid_rotation}, the orbital period is too long for binary circularization and is much longer than the rotational periods, suggesting that the stars' rotation periods are unaffected by J1921's binarity \citep{mayor_binary_cutoff}.

The two stars can be classified as either RS Canum Venaticorum (RS CVn) or BY Draconis (BY Dra) type variables.
The Variable Stars Index \citep[VSX;][]{vsx_2006} lists the requirements for an RS CVn-type variable as (1) binary components that are of late F to late K spectral type (usually giants), (2) the presence of strong Ca \textit{II} H and K emission lines, (3) the presence of radio and X-ray emission, and (4) a sinusoidal light curve outside the eclipses. The criteria for BY Dra-type variable stars exclude indicators of chromospheric activity, where the only requirements are that they are emission-line dwarfs (dKe-dMe spectral type) and have quasi-periodic light changes with periods from a fraction of a day to 120 days and amplitudes from several hundredths to $0.5$ mag in \textit{V}. In the first \textit{Catalog of Chromospherically Active Binaries} \citep{CABs_1}, they note that the difference between these two variable types is "useful mainly from an historical viewpoint" as many chromospherically active stars can be classified as either type. We will show that J1921 satisfies the criteria for chromospheric activity and the rotational variable types listed above.

We perform a frequency analysis of the ASAS-SN and \textit{TESS} light curves in \S\ref{sec:lc}. 
We use its spectrum and spectral energy ditribution (SED) to characterize the primary star in \S\ref{sec:pri}.
In \S\ref{sec:model}, we use \textsc{phoebe} \citep{phoebe_2020_v23} to model the orbit and the ratios of stellar parameters. 
We end with a brief summary in \S\ref{sec:sum}.
Throughout our analysis, we use a distance of $315 \pm 2$ pc from \cite{Gaia_dist}. The Galactic reddening in the direction of J1921 is $E(B-V) = 0.047$ mag \citep{schlafly_finkbeiner_exctinction}, but we expect most of the dust to be behind J1921, due to its proximity.

\begin{figure}
    \includegraphics[width=0.9\columnwidth]{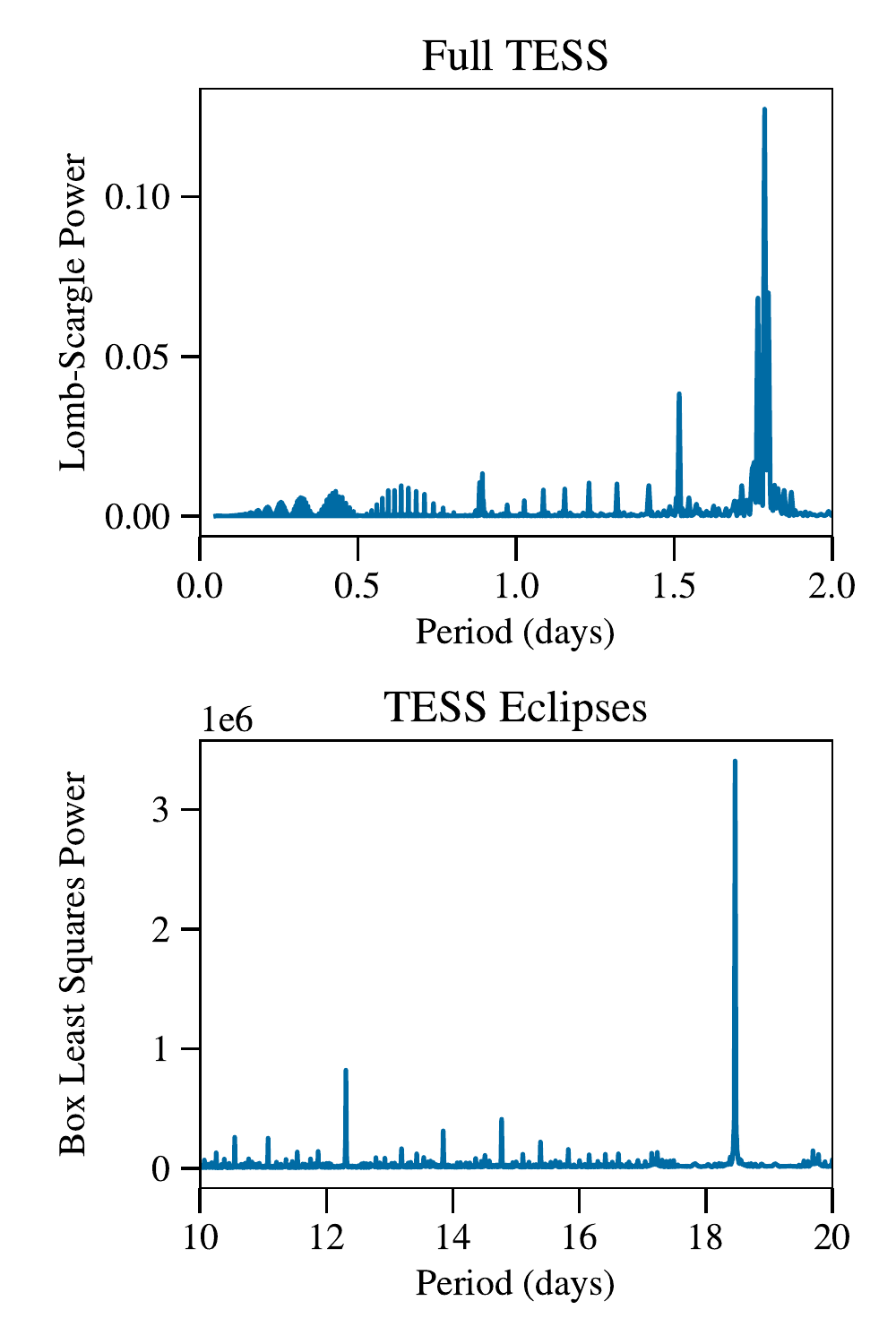}
    \caption{\textit{Top Panel:} The Lomb-Scargle periodogram for the \textit{TESS} data shown at short periods to search for the rotational signals. \textit{Bottom Panel:} The Box Least Square periodogram for the masked \textit{TESS} data used to search for the longer orbital period.}
    \label{fig:periodogram_tess}
\end{figure}

\section{Discovery and Light Curves}
\label{sec:lc}

In this section we will analyze the general properties of the light curves observed by ASAS-SN and \textit{TESS}. Figure \ref{fig:ASAS_lc} shows the ASAS-SN g- and V-band light curves dating back to UT 2013-02-21.61. The eclipses captured by these data are shown using big circles, the last of which motivated us to investigate this source more thoroughly. 
There are two, slightly dimmed points on UT 2018-04-23.49 and UT 2018-09-01.38. The former is outside of both the primary and secondary eclipse's full width at half max (FWHM) while the the latter coincides with a primary eclipse.

\begin{figure}
    \includegraphics[width=0.9\columnwidth]{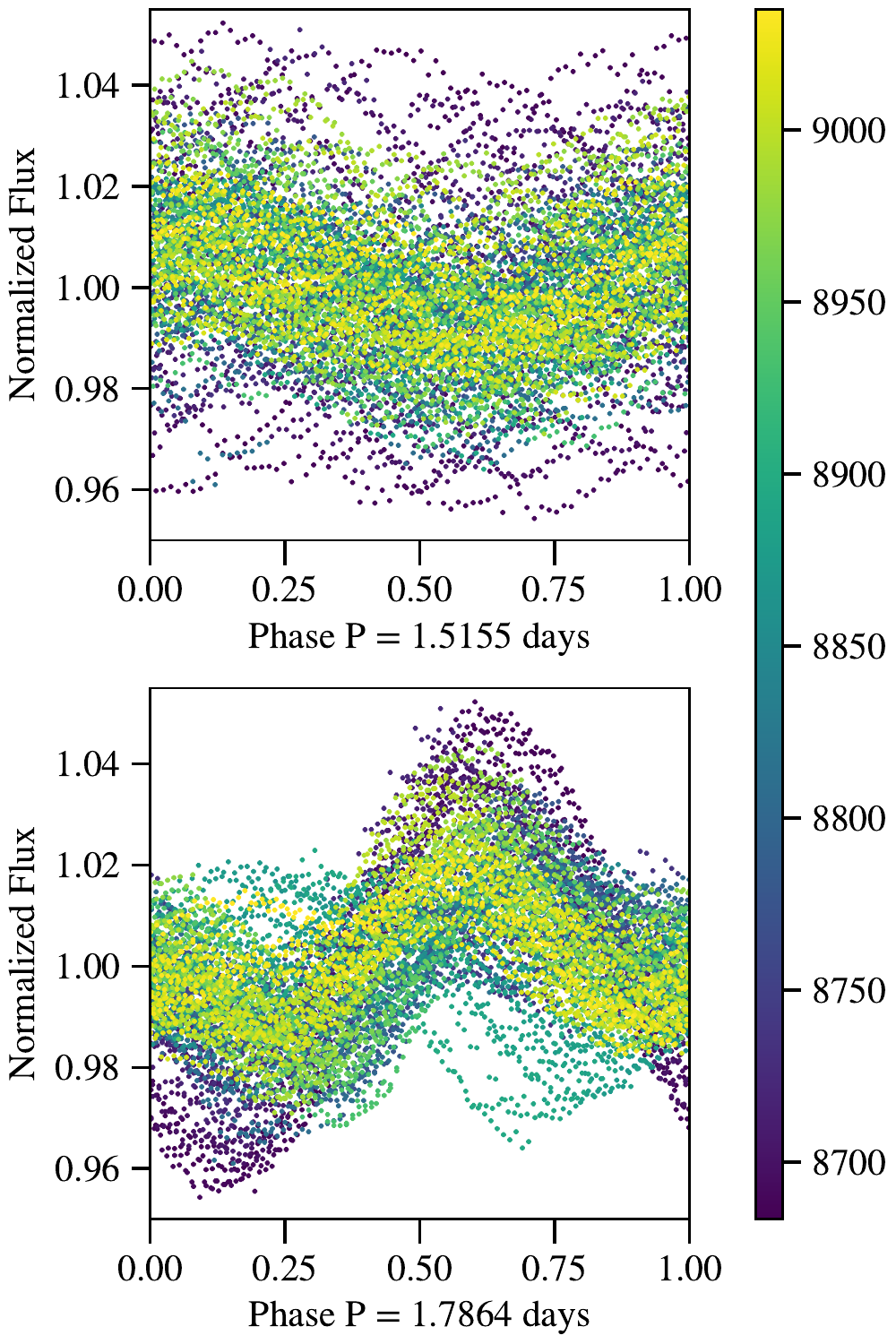}
    \caption{The \textit{TESS} data phased to the rotation periods found in the Lomb-Scargle periodogram. The color is coded by the epoch of the observations to show how the light curve changes with time. The eclipses have been masked.}
    \label{fig:tess_phased}
\end{figure}

A Lomb-Scargle \citep{lombscargle} periodogram of the ASAS-SN V-band data is dominated by diurnal aliasing, causing J1921 to be rejected as a rotational variable source in \cite{tharindu_variable_1}. A Lomb-Scargle periodogram of the g-band data shows a clear peak at $\sim 1.79$ days that corresponds to one of the the rotational signals we find below. This period was also detected by the Zwicky Transient Facility (ZTF) \citep{Zwicky_2020}. The second rotational period at $\sim 1.52$ days was not detected in the ground-based data. In addition to the eclipses, there is a slow, secular rise and fall in the V-band flux which we believe is real but do not discuss further in this paper.

Figure \ref{fig:Tess_lc} shows 11 sectors of \textit{TESS} observations from UT 2019-07-18.86 to UT 2019-11-2.17 normalized so that the median flux is 1. 
The light curve was produced using the adaption of the ASAS-SN difference imaging pipeline for \textit{TESS} data described in \cite{Vallely2020}.
Each sector's differential light curve was normalized to have a mean flux equal to the \textit{TESS} Input Catalog source ($T = 12.282$ mag).
The nature of the source is now clear -- both primary and secondary eclipses of the eccentric orbit are seen against a baseline of quasiperiodic variability on a time scale much shorter than the orbital period.

Figure \ref{fig:periodogram_tess} shows the Lomb-Scargle periodogram \citep{lombscargle} of the \textit{TESS} data. There are two significant peaks, with a strong signal at $\sim 1.79$ days and a weaker one at $\sim 1.52$ days. Figure \ref{fig:tess_phased} shows the \textit{TESS} data folded to these periods and excluding the eclipses. Our hypothesis is that these two periods are due to the rotational modulation of spots on the primary and secondary stars. There are a series of smaller amplitude peaks that are due to the eclipses, but the Lomb-Scargle periodogram does not clearly show a signal at the orbital period.

The Lomb-Scargle algorithm works poorly for systems with narrow eclipses, so we used the box least squares algorithm \citep{2002A&A...391..369K} to find the orbital period. 
We masked the out-of-eclipse data by replacing any normalized flux $> 0.964$ with $1$, leaving a flat light curve punctuated by the eclipses. 
This prevents the box least squares algorithm from detecting the rotational signals.
The lower panel in Figure \ref{fig:periodogram_tess} shows the resulting periodogram with a peak at the eclipsing binary period of $P_{\rm orb} = 18.46199$ days. Figure \ref{fig:Tess_lc} shows the light curve in the top panel folded to this period. The asymmetry of the eclipse phases means that the binary orbit is very elliptical. The eclipses at their FWHM only cover 1.05\% of the total phase, which explains why ASAS-SN only observed a few dimming events, even with its $\sim 1$ day g-band cadence.

\section{Characterizing the Primary}
\label{sec:pri}

\setlength\extrarowheight{1pt}
\begin{table}
\centering
 \caption{Photometry of J1921.}
 \label{tab:photometry}
 \begin{tabular}{lccc}
 \hline
Survey & Band &  Mag &  Uncertainty \\
\hline
GALEX & NUV*  &  18.7752 &     0.0510\\
\hline
SDSS & u & 15.447 & 0.005 \\
\hline
APASS &   i'   & 12.540 &    $--$** \\
APASS & r'    &  12.808 &     0.008\\
APASS  & V &  13.127 &     0.012\\
APASS  & g' &    13.443 &     0.040 \\
APASS  & B & 13.897 &     0.034\\
\hline
2MASS & $\text{K}_s$  &   10.806 &     0.019 \\
2MASS & H  &   10.939 &     0.021\\
2MASS & J  &   11.388 &     0.020\\
\hline
WISE & W2  &   10.774 &     0.020 \\
WISE & W1  &   10.753 &     0.023 \\
\hline
\multicolumn{4}{l}{\footnotesize{* Used as an upper bound}} \\
\multicolumn{4}{l}{\footnotesize{** No error was given}} \\
 \end{tabular}
\end{table}

\begin{figure*}
\includegraphics[width=\textwidth]{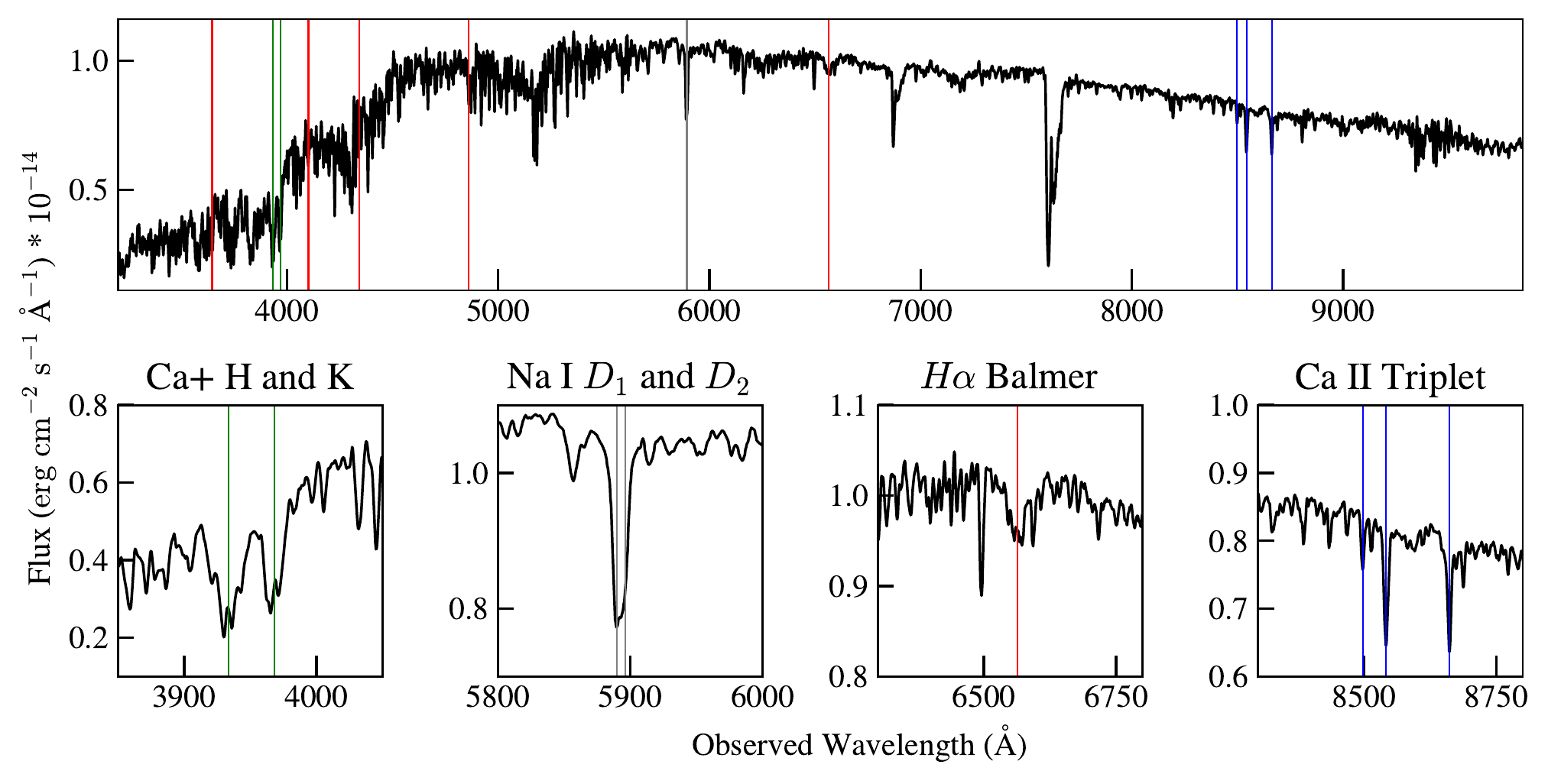}
    \caption{The LBT/MODS spectrum of J1921. Lines mark the Ca+ H and K (green), Na $D_1$ and $D_2$ (gray), H$\alpha$ absorption (red), and the Ca II triplet (blue) in the lower panels from left to right.}
    \label{fig:LBT_spec}
\end{figure*}

We can characterize the primary star by using its spectrum and SED. Neither should be strongly effected by the secondary because it is significantly dimmer than the primary. 
We obtained a spectrum of ASASSN-V J192114.84+624950.8 on Dec 21, 2019 using the Multi-Object Double Spectrographs mounted on the twin 8.4m Large Binocular Telescope \citep[MODS;][]{PoggeMODS}.
This spectrum was reduced using a standard combination of the \textsc{modsccdred}\footnote{\scriptsize{\url{http://www.astronomy.ohio-state.edu/MODS/Software/modsCCDRed/}}} \textsc{python} package, and the \textsc{modsidl} pipeline\footnote{\scriptsize{\url{http://www.astronomy.ohio-state.edu/MODS/Software/modsIDL/}}}.
We show the spectrum in Figure \ref{fig:LBT_spec}. The spectrum is that of a late-G or early-K type dwarf, consistent with the temperature and luminosity reported in Gaia DR2 of $4995^{+220}_{-70}$ K and $0.603 \pm 0.008$ $L_{\sun}$ \citep{Gaia_2018b}.

J1921 shows evidence of chromospheric activity, a phenomena produced by rapid rotation and thick convection zones \citep{1984ApJ...279..763N}.
A thorough list of spectroscopic indicators of chromospheric activity can be found in Section 3 of \cite{Zhang_chromo}. Of these indicators, we see evidence for activity in the Na \textit{I} $D_1$ and $D_2$, Ca \textit{II} infrared triplet, H$\alpha$ Balmer, and Ca \textit{II} H \& K lines. These features are shown in the panels below the full spectrum in Figure \ref{fig:LBT_spec}.
The Frauenhofer Ca \textit{II} H and K feature is consistent in shape and depth with previously identified chromospherically active binaries, particularly the examples shown in Appendix B of \cite{1990ApJS...72..191S}. 
In addition to the spectral indicators of activity, J1921 is an X-ray source (2RXS J192114.7+624951) in the second ROSAT survey \citep{2016A&A...588A.103B}. 
Based on the orbit and the stellar activity, we classify J1921 as a chromospherically active, rotational variable binary system.
\cite{Zwicky_2020} classified J1921 as a BY Draconis variable due to its $P=1.79$ day rotational modulation.

Table \ref{tab:photometry} lists the photometry we used to model the primary's spectral energy distribution (SED). Because the source is so bright, many standard sources of photometry are saturated. Here we use the AllWISE catalog \citep{ALLWISE}, 2MASS \citep{2MASS_catalog}, APASS \citep{2015AAS...22533616H} and the SDSS u-band \citep{sdss_12}.
There is a GALEX \citep{2017ApJS..230...24B} NUV detection that we use as an upper limit because of the evidence for chromospheric activity.

We fit the SED with DUSTY \citep{DUSTY, DUSTY_2} inside a Markov Chain Monte Carlo (MCMC) wrapper \citep{mcmc_dusty}. We use the Gaia DR2 \citep{Gaia_dist} distance of $315$ pc with no foreground extinction. 
For our stellar atmospheres, we used the models in \cite{2003IAUS..210P.A20C} and assumed minimal flux errors of $10\%$ to account for any systematic issues. This leads to a best fit model with $\chi^2 = 8.4$ that has a temperature of $T_{e} = 4930$ K and a total luminosity of $L = 0.62~L_{\sun}$. This is consistent with our spectral classification and the Gaia DR2 \citep{Gaia_2018b} temperature estimate. 
Given the results in \S\ref{sec:model} for the radius and temperature ratios of the two components, we roughly modeled the SED as two stars and find that the SED is well fit by a $T_{e,1} = 5050~K$, $L_1 = 0.48 ~ L_{\sun}$ primary combined with a $T_{e,2} = 4190~K$, $L_2 = 0.12~L_{\sun}$ secondary. The SEDs both of the component stars and their sum are shown in Figure~\ref{fig:sed_kochanek}.

\afterpage{\

\begin{figure}
\includegraphics[width=\columnwidth]{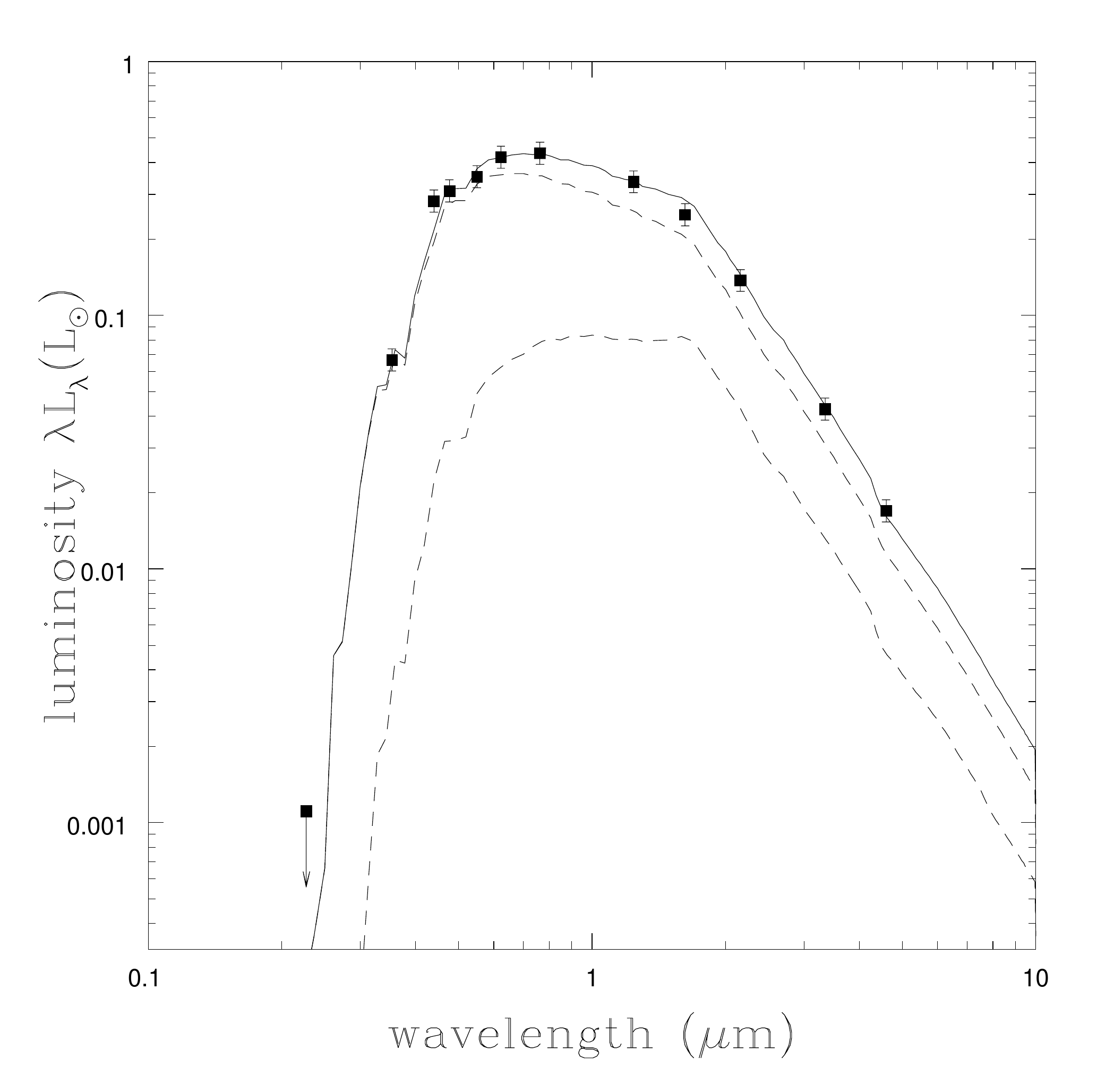}
    \caption{The SED of the system shown with models for the primary and secondary stars (dashed) and their sum (solid).}
    \label{fig:sed_kochanek}
\end{figure}

\begin{figure}
    \includegraphics[width=\columnwidth]{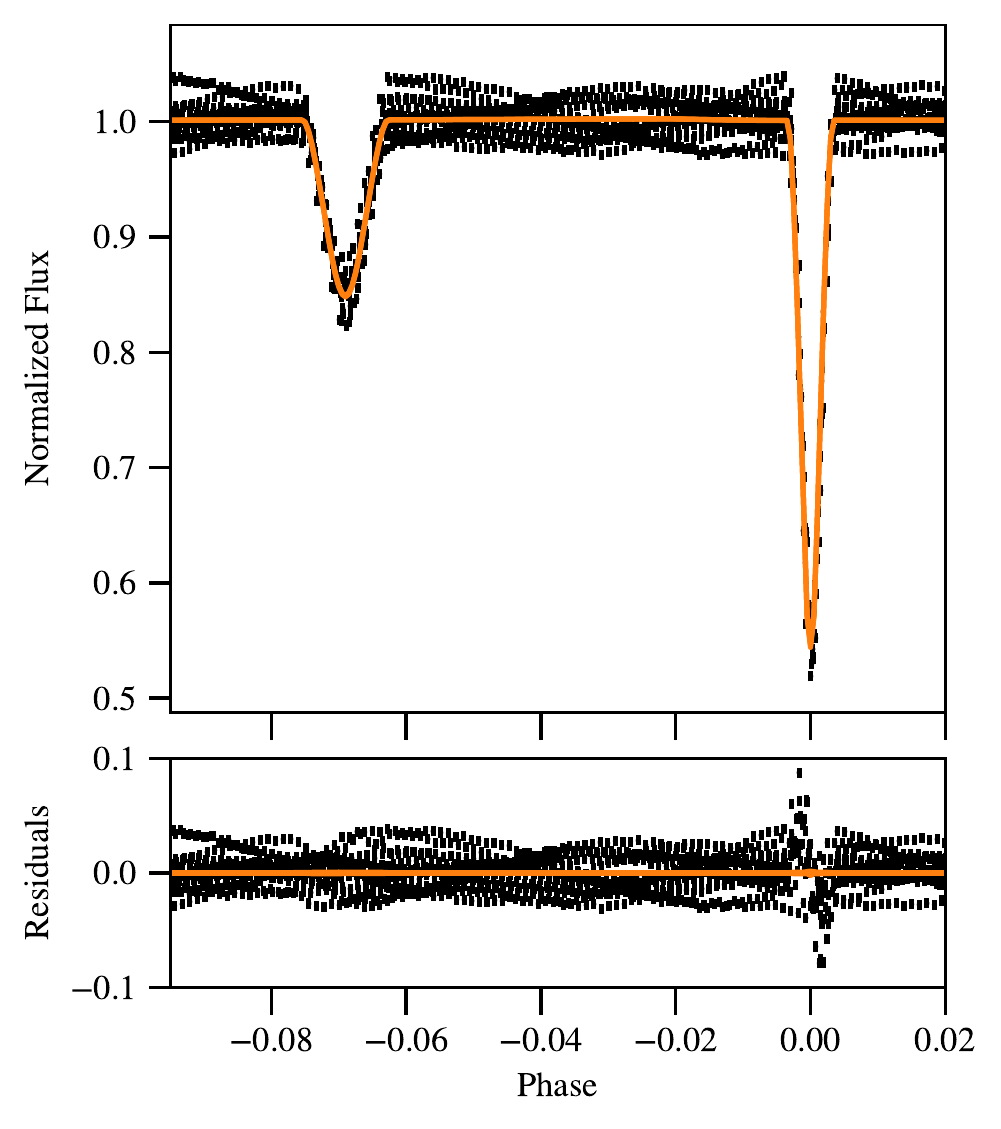}
    \caption{The top panel shows the \textit{TESS} data with 100 light curves sampled within $1\sigma$ of our posterior distribution (see Figure \ref{fig:mcmc}). The bottom panel shows the flux residuals. The errors of the \textit{TESS} data are artificially inflated by a factor of 5 to prevent over-fitting.}
    \label{fig:model}
\end{figure}

\clearpage
}

\afterpage{
\begin{figure*}
    \centering
    \includegraphics[width=0.9\textwidth]{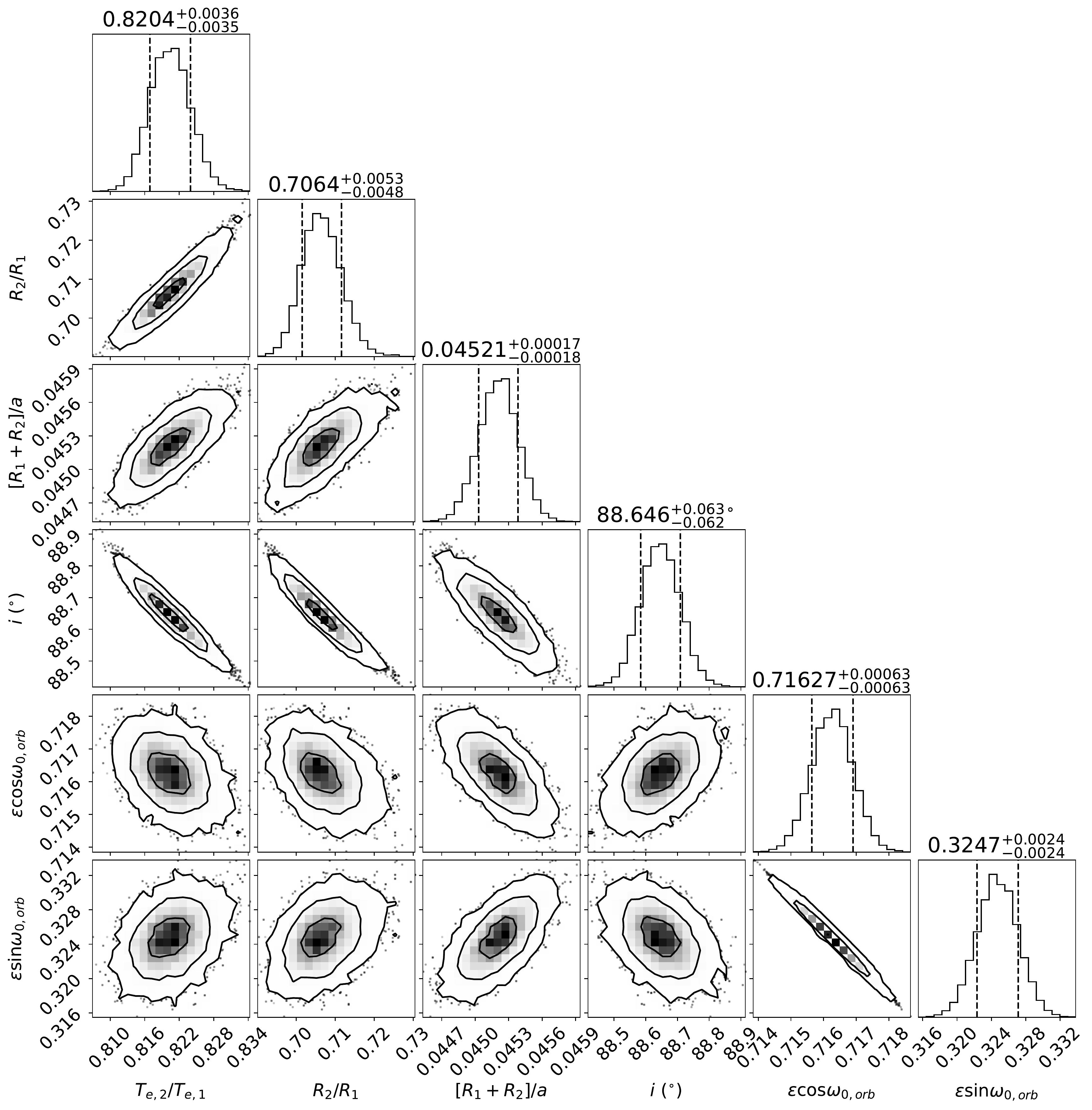}
    \caption{The posterior distribution of the binary and stellar parameters optimised in the \textsc{phoebe} fits to the \textit{TESS} data in \S\ref{sec:model}}
    \label{fig:mcmc}
\end{figure*}
\clearpage
}

\section{Eclipse Modeling}
\label{sec:model}

We modeled the \textit{TESS} light curve with the PHysics Of Eclipsing BinariEs \citep[\textsc{phoebe};][]{Pr_a_2005, Pr_a_2016, Horvat_2018, phoebe_2020_v23} \textsc{python} code. We use a fixed period of $P_{\rm orb}=18.46199$ days. We artificially inflate the \textit{TESS} photometric errors by a factor of 5 because the they are dominated by systematics.
Even without radial velocities, we can constrain the effective temperatures, radii, eccentricity, argument of periapsis, and inclination of the system. We use the \textsc{ellc} \citep{ellc} package to model our light curves in the \textsc{phoebe} architecture.

We begin our analysis by roughly estimating the ratio of the effective temperatures. We can approximate this from the \textit{TESS} light curve using \citep[eq. 7.11]{intro_to_textbook}
\begin{equation}
    \frac{B_{0} - B_{1}}{B_{0} - B_{2}} = \left(\frac{T_{e,2}}{T_{e,1}}\right) ^4
\end{equation}
where $B_{0}$ is the out of eclipse flux, $B_{i}$ are the flux minima during eclipse, and $T_{e,i}$ are the component effective temperatures. We find $T_{e,2}/T_{e,1} \sim 0.76$.

We then use \textsc{phoebe}'s Gaussian eclipse model to estimate the geometry of the light curve. Using this estimate we can approximate the eccentricity ($e \sim 0.783$) and argument of periapsis ($\omega_0 \sim 22.94^{\circ}$).
Given these initial estimates, we optimized the model using the Nelder-Mead simplex algorithm \citep{nelder_mead} to estimate the ratio of effective temperatures ($T_{e,2}/T_{e,1}$), the ratio of radii ($R_2/R_1$), the ratio of the summed radii to the semi-major axis ($\left[ R_1 + R_2 \right]/a$), inclination ($i$), eccentricity ($e$), and argument of periapsis ($\omega_0$). We keep the the effective temperature ($T_{e,1}$) and radius ($R_1$) of the primary constrained by the single star values found in \S\ref{sec:pri} for this optimization since these fits cannot determine either quantity.

\begin{figure}
    \centering
    \includegraphics[width=\columnwidth]{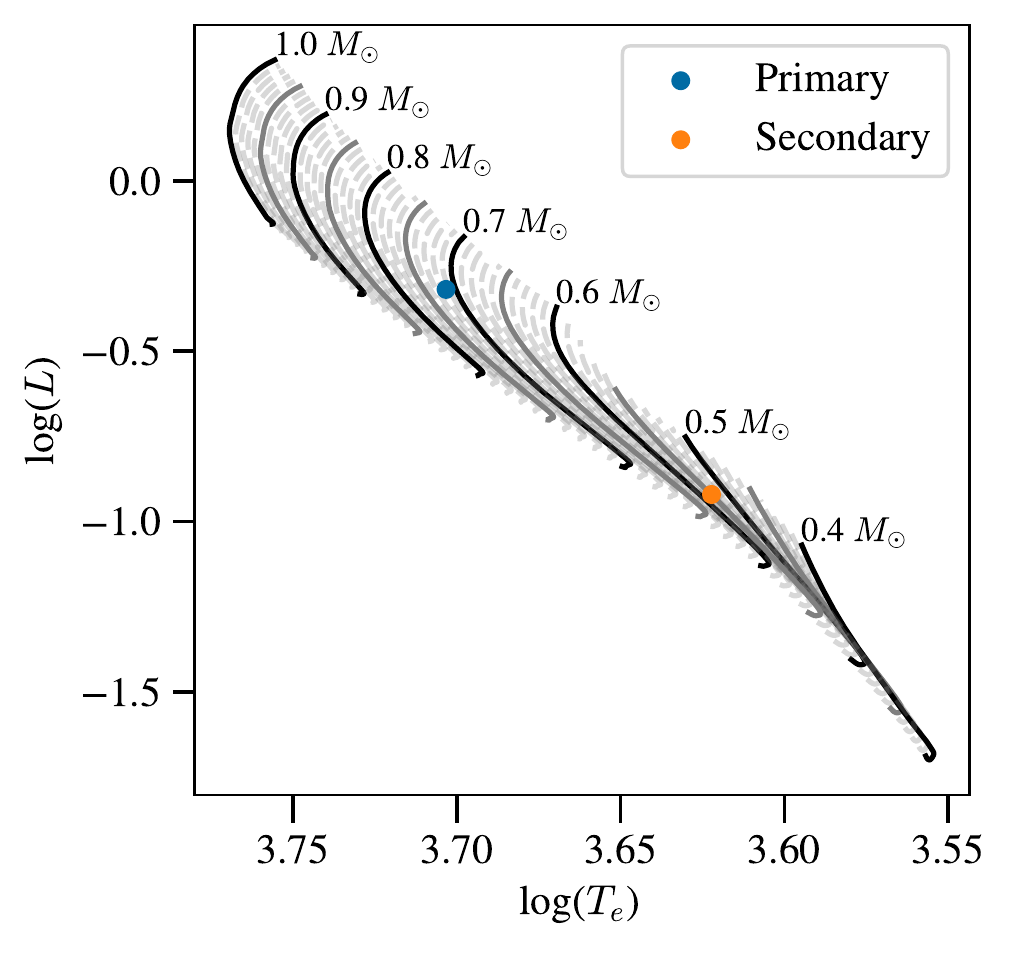}
    \caption{The primary and secondary temperature and luminosities are shown over a range of MIST evolutionary tracks. The isochrones span $0.4~M_{\sun}$ to $1~M_{\sun}$ with every multiple of $0.05 M_{\sun}$ shown as a solid line. The MIST models are evolved over the main-sequence with solar metallicity.}
    \label{fig:MIST}
\end{figure}

Finally, we estimated the uncertainties using the \textsc{emcee} \citep{emcee} Markov Chain Monte Carlo package integrated into \textsc{phoebe}.
Based on early fits and the SED fit at the end of \S\ref{sec:pri}, we fix $q=0.3$, $T_{e,1}=5050~K$, and $R_1 = 0.907~R_{\sun}$ for our analysis.
We used Gaussian priors centered around the Nelder-Mead results and marginalize over, but do not directly fit, the primary's passband luminosity ($L_{pb}$).
Figure \ref{fig:model} shows 100 light curves sampled within $1\sigma$ of our posterior distribution. The full distribution is shown in Figure \ref{fig:mcmc} with the parameter values listed in the third column of table \ref{tab:mcmc_params}. For this model, periastron would occur at phase $-0.017$. We do not see any obvious residual anomalies here.

\begin{table}
    \centering
	\label{tab:mcmc_params}
	\caption{The parameters fit in our final MCMC analysis. Variables that are ratios must be greater than zero. The posterior uncertainties are rounded to the largest significant figure.}
	\begin{tabular}{lcc}
		Variable & Prior $\pm 1\sigma$ & Posterior\\
		\hline
		$T_{e,2}/T_{e,1}$                     & $0.83 \pm 0.05$ &  $0.820 \pm 0.004$\\[2pt]
		$R_2/R_1$                         & $0.73 \pm 0.10$ &  $0.706 \pm 0.005$\\[2pt]
		$\left( R_1 + R_2 \right)/a$                &$0.05\pm0.10$ &  $0.0456 \pm 0.0002$\\[2pt]
		Inclination ($i$)                  & $88.44 \pm 0.50^{\circ}$  & $88.65 \pm 0.06^{\circ}$\\[2pt]
		$e \cos{\omega_0}$                  & $0.71 \pm 0.10$ & $0.7163 \pm 0.0006$\\[2pt]
		$e \sin{\omega_0}$                  & $0.33 \pm 0.10$  & $0.325 \pm 0.002$\\[2pt] 
		$e$* &$--$& $0.7864 \pm 0.0004$ \\[2pt]
		\hline
		\multicolumn{3}{l}{* \footnotesize{Calculated from posterior distribution}} \\
	\end{tabular}
\end{table}

Our model is incomplete until a radial velocity curve is observed. However, using the model from \S\ref{sec:model}, we can compare the effective temperatures and radii to MIST evolutionary tracks \citep{MIST_0, MIST_1, MESA_1, MESA_2, MESA_3}. Figure \ref{fig:MIST} shows a range of stellar models with solar metallicity evolving on or near the main sequence. The luminosities and temperatures of the stars are consistent with masses $M_1 \sim 0.71 ~ M_{\sun}$ and $M_2 \sim 0.55 ~ M_{\sun}$. This is consistent with our spectral analysis in \S\ref{sec:pri}. 



\section{Summary}
\label{sec:sum}

During the ongiong All-Sky Automated Survey for SuperNovae \citep[ASAS-SN;][]{2014ApJ...788...48S, 2017PASP..129j4502K} survey, we discovered the variability of ASASSN-V J192114.84+624950.8 as a $\Delta g = 0.88$ mag dip in brightness.
Then, using \textit{TESS} \citep{tess_instrument_paper} photometry, we found the source to be a highly eccentric, eclipsing binary with rotational variation.
A Lomb-Scargle periodogram of the light curve shows the periods of rotation at $P=1.52$ and $P=1.79$ days.
A Box Least Squares periodogram of the light curve, modified to mask the rotational variability, reveals the orbital period to be $P_{\rm orb} = 18.46$ days.

In order to characterize the system, a spectrum of J1921 was taken using the Multi-Object Double Spectrographs mounted on the twin 8.4m Large Binocular Telescope \citep[MODS;][]{PoggeMODS}. The spectrum is that of a late-G or early-K type dwarf with indicators of chromospheric activity.
We fit the spectral energy distribution as a sum of the two component stars and find a best fit model with $L_1 = 0.48~L_{\sun}$, $T_1= 5050~K$, $L_2 = 0.12~L_{\sun}$, and $T_{2} = 4190~K$. 

We conclude by using the \textsc{phoebe} \citep{Pr_a_2005, Pr_a_2016, Horvat_2018, phoebe_2020_v23} and \textsc{ellc} \citep{ellc} packages to model the light curve eclipses. After some optimization, we determine the errors of the orbital parameters with the \textsc{emcee} \citep{emcee} Markov Chain Monte Carlo package.
We find the eccentricity and inclination to be $e = 0.79$ and $i=88.65^{\circ}$. A full list of parameters and their errors can be found in Table \ref{tab:mcmc_params}.

This system, due to it's proximity and peculiar orbit, may be interesting for the outstanding discrepancy between theoretically predicted and actual radii of late type stars \citep{mdwarf_problem}. A thorough analysis of the radial velocity curve, multi-band eclipse observations, and the spectroscopic properties of the stars should yield a complete solution for the masses, radii, luminosities, and temperatures of the two stars, allowing comparisons to stellar models similar to \cite{mdwarf_problem}.

\section*{Acknowledgements}

PJV is supported by the National Science Foundation Graduate Research Fellowship Program Under Grant No. DGE-1343012. TJ is supported by the Ohio State University Presidential Fellowship. CSK and KZS are supported by NSF grants AST-1814440 and AST-1908570. Support for TW-SH was provided by NASA through the NASA Hubble Fellowship grant HST-HF2-51458.001-A awarded by the Space Telescope Science Institute, which is operated by the Association of Universities for Research in Astronomy, Inc., for NASA, under contract NAS5-265. 
B.J.S. is supported by NSF grants AST-1907570, AST-1908952, AST-1920392, and AST-1911074. 

We thank Las Cumbres Observatory and its staff for their continued support of ASAS-SN.
ASAS-SN is funded in part by the Gordon and Betty Moore Foundation through grants GBMF5490 and GBMF10501 to the Ohio State University, NSF grant AST-1908570, the Mt. Cuba Astronomical Foundation, the Center for Cosmology and AstroParticle Physics (CCAPP) at OSU, the Chinese Academy of Sciences South America Center for Astronomy (CAS-SACA), and the Villum Fonden (Denmark). Development of ASAS-SN has been supported by NSF grant AST-0908816, the Center for Cosmology and AstroParticle Physics at the Ohio State University, the Mt. Cuba Astronomical Foundation, and by George Skestos. 

This paper includes data collected by the \textit{TESS} mission, which are publicly available from the Mikulski Archive for Space Telescopes (MAST). Funding for the \textit{TESS} mission is provided by NASA's Science Mission directorate.

The LBT is an international collaboration among institutions in the United States, Italy and Germany.
LBT Corporation partners are: The Ohio State University, and The Research Corporation, on behalf of The University of Notre Dame, University of Minnesota and University of Virginia; The University of Arizona on behalf of the Arizona university system; Istituto Nazionale di Astrofisica, Italy; LBT Beteiligungsgesellschaft, Germany, representing the Max-Planck Society, the Astrophysical Institute Potsdam, and Heidelberg University.

This publication makes use of data products from the Wide-field Infrared Survey Explorer, which is a joint project of the University of California, Los Angeles, and the Jet Propulsion Laboratory/California Institute of Technology, funded by the National Aeronautics and Space Administration.

This research has made use of the VizieR catalogue accesstool, CDS, Strasbourg, France. 
This research also made use of Astropy, a community-developed core Python package for Astronomy \citep{astropy:2018}.

\section*{Data Availability}
Except for the LBT spectrum, all data used in this paper are publically available. The spectrum will be shared on reasonable request to the corresponding author.

\newpage




\bibliographystyle{mnras}
\bibliography{paper_text}

\begin{thebibliography}{}
\makeatletter
\relax
\def\mn@urlcharsother{\let\do\@makeother \do\$\do\&\do\#\do\^\do\_\do\%\do\~}
\def\mn@doi{\begingroup\mn@urlcharsother \@ifnextchar [ {\mn@doi@}
  {\mn@doi@[]}}
\def\mn@doi@[#1]#2{\def\@tempa{#1}\ifx\@tempa\@empty \href
  {http://dx.doi.org/#2} {doi:#2}\else \href {http://dx.doi.org/#2} {#1}\fi
  \endgroup}
\def\mn@eprint#1#2{\mn@eprint@#1:#2::\@nil}
\def\mn@eprint@arXiv#1{\href {http://arxiv.org/abs/#1} {{\tt arXiv:#1}}}
\def\mn@eprint@dblp#1{\href {http://dblp.uni-trier.de/rec/bibtex/#1.xml}
  {dblp:#1}}
\def\mn@eprint@#1:#2:#3:#4\@nil{\def\@tempa {#1}\def\@tempb {#2}\def\@tempc
  {#3}\ifx \@tempc \@empty \let \@tempc \@tempb \let \@tempb \@tempa \fi \ifx
  \@tempb \@empty \def\@tempb {arXiv}\fi \@ifundefined
  {mn@eprint@\@tempb}{\@tempb:\@tempc}{\expandafter \expandafter \csname
  mn@eprint@\@tempb\endcsname \expandafter{\@tempc}}}

\bibitem[\protect\citeauthoryear{{Adams} \& {Kochanek}}{{Adams} \&
  {Kochanek}}{2015}]{mcmc_dusty}
{Adams} S.~M.,  {Kochanek} C.~S.,  2015, \mn@doi [\mnras]
  {10.1093/mnras/stv1409}, \href
  {https://ui.adsabs.harvard.edu/abs/2015MNRAS.452.2195A} {452, 2195}

\bibitem[\protect\citeauthoryear{{Alam} et~al.,}{{Alam} et~al.}{2015}]{sdss_12}
{Alam} S.,  et~al., 2015, \mn@doi [\apjs] {10.1088/0067-0049/219/1/12}, \href
  {https://ui.adsabs.harvard.edu/abs/2015ApJS..219...12A} {219, 12}

\bibitem[\protect\citeauthoryear{{Astropy Collaboration} et~al.,}{{Astropy
  Collaboration} et~al.}{2018}]{astropy:2018}
{Astropy Collaboration} et~al., 2018, \mn@doi [aj] {10.3847/1538-3881/aabc4f},
  \href {https://ui.adsabs.harvard.edu/abs/2018AJ....156..123A} {156, 123}

\bibitem[\protect\citeauthoryear{{Bailer-Jones}, {Rybizki}, {Fouesneau},
  {Mantelet}  \& {Andrae}}{{Bailer-Jones} et~al.}{2018}]{Gaia_dist}
{Bailer-Jones} C.~A.~L.,  {Rybizki} J.,  {Fouesneau} M.,  {Mantelet} G.,
  {Andrae} R.,  2018, \mn@doi [\aj] {10.3847/1538-3881/aacb21}, \href
  {https://ui.adsabs.harvard.edu/abs/2018AJ....156...58B} {156, 58}

\bibitem[\protect\citeauthoryear{{Bianchi}, {Shiao}  \& {Thilker}}{{Bianchi}
  et~al.}{2017}]{2017ApJS..230...24B}
{Bianchi} L.,  {Shiao} B.,   {Thilker} D.,  2017, \mn@doi [\apjs]
  {10.3847/1538-4365/aa7053}, \href
  {https://ui.adsabs.harvard.edu/abs/2017ApJS..230...24B} {230, 24}

\bibitem[\protect\citeauthoryear{{Boller}, {Freyberg}, {Tr{\"u}mper}, {Haberl},
  {Voges}  \& {Nandra}}{{Boller} et~al.}{2016}]{2016A&A...588A.103B}
{Boller} T.,  {Freyberg} M.~J.,  {Tr{\"u}mper} J.,  {Haberl} F.,  {Voges} W.,
  {Nandra} K.,  2016, \mn@doi [\aap] {10.1051/0004-6361/201525648}, \href
  {https://ui.adsabs.harvard.edu/abs/2016A&A...588A.103B} {588, A103}

\bibitem[\protect\citeauthoryear{{Carroll} \& {Ostlie}}{{Carroll} \&
  {Ostlie}}{2006}]{intro_to_textbook}
{Carroll} B.~W.,  {Ostlie} D.~A.,  2006, {An introduction to modern
  astrophysics and cosmology}.
Pearson Addison-Wesley

\bibitem[\protect\citeauthoryear{{Castelli} \& {Kurucz}}{{Castelli} \&
  {Kurucz}}{2003}]{2003IAUS..210P.A20C}
{Castelli} F.,  {Kurucz} R.~L.,  2003, in {Piskunov} N.,  {Weiss} W.~W.,
  {Gray} D.~F.,  eds, ~ Vol. 210, Modelling of Stellar Atmospheres. p.~A20
  (\mn@eprint {arXiv} {astro-ph/0405087})

\bibitem[\protect\citeauthoryear{{Chen}, {Wang}, {Deng}, {de Grijs}, {Yang}  \&
  {Tian}}{{Chen} et~al.}{2020}]{Zwicky_2020}
{Chen} X.,  {Wang} S.,  {Deng} L.,  {de Grijs} R.,  {Yang} M.,   {Tian} H.,
  2020, \mn@doi [\apjs] {10.3847/1538-4365/ab9cae}, \href
  {https://ui.adsabs.harvard.edu/abs/2020ApJS..249...18C} {249, 18}

\bibitem[\protect\citeauthoryear{{Choi}, {Dotter}, {Conroy}, {Cantiello},
  {Paxton}  \& {Johnson}}{{Choi} et~al.}{2016}]{MIST_1}
{Choi} J.,  {Dotter} A.,  {Conroy} C.,  {Cantiello} M.,  {Paxton} B.,
  {Johnson} B.~D.,  2016, \mn@doi [\apj] {10.3847/0004-637X/823/2/102}, \href
  {https://ui.adsabs.harvard.edu/abs/2016ApJ...823..102C} {823, 102}

\bibitem[\protect\citeauthoryear{{Conroy} et~al.,}{{Conroy}
  et~al.}{2020}]{phoebe_2020_v23}
{Conroy} K.~E.,  et~al., 2020, \mn@doi [\apjs] {10.3847/1538-4365/abb4e2},
  \href {https://ui.adsabs.harvard.edu/abs/2020ApJS..250...34C} {250, 34}

\bibitem[\protect\citeauthoryear{{Cutri} \& {et al.}}{{Cutri} \& {et
  al.}}{2014}]{ALLWISE}
{Cutri} R.~M.,  {et al.} 2014, VizieR Online Data Catalog, \href
  {https://ui.adsabs.harvard.edu/abs/2014yCat.2328....0C} {p. II/328}

\bibitem[\protect\citeauthoryear{{Cutri} et~al.,}{{Cutri}
  et~al.}{2003}]{2MASS_catalog}
{Cutri} R.~M.,  et~al., 2003, VizieR Online Data Catalog, \href
  {https://ui.adsabs.harvard.edu/abs/2003yCat.2246....0C} {p. II/246}

\bibitem[\protect\citeauthoryear{{Dotter}}{{Dotter}}{2016}]{MIST_0}
{Dotter} A.,  2016, \mn@doi [\apjs] {10.3847/0067-0049/222/1/8}, \href
  {https://ui.adsabs.harvard.edu/abs/2016ApJS..222....8D} {222, 8}

\bibitem[\protect\citeauthoryear{{Elitzur} \& {Ivezi{\'c}}}{{Elitzur} \&
  {Ivezi{\'c}}}{2001}]{DUSTY_2}
{Elitzur} M.,  {Ivezi{\'c}} {\v{Z}}.,  2001, \mn@doi [\mnras]
  {10.1046/j.1365-8711.2001.04706.x}, \href
  {https://ui.adsabs.harvard.edu/abs/2001MNRAS.327..403E} {327, 403}

\bibitem[\protect\citeauthoryear{{Foreman-Mackey}, {Hogg}, {Lang}  \&
  {Goodman}}{{Foreman-Mackey} et~al.}{2013}]{emcee}
{Foreman-Mackey} D.,  {Hogg} D.~W.,  {Lang} D.,   {Goodman} J.,  2013, \mn@doi
  [\pasp] {10.1086/670067}, \href
  {https://ui.adsabs.harvard.edu/abs/2013PASP..125..306F} {125, 306}

\bibitem[\protect\citeauthoryear{{Gaia Collaboration} et~al.,}{{Gaia
  Collaboration} et~al.}{2018}]{Gaia_2018b}
{Gaia Collaboration} et~al., 2018, \mn@doi [\aap]
  {10.1051/0004-6361/201833051}, \href
  {https://ui.adsabs.harvard.edu/abs/2018A&A...616A...1G} {616, A1}

\bibitem[\protect\citeauthoryear{Gao \& Han}{Gao \& Han}{2012}]{nelder_mead}
Gao F.,  Han L.,  2012, \mn@doi [Computational Optimization and Applications]
  {10.1007/s10589-010-9329-3}, 51, 259

\bibitem[\protect\citeauthoryear{{Henden}, {Levine}, {Terrell}  \&
  {Welch}}{{Henden} et~al.}{2015}]{2015AAS...22533616H}
{Henden} A.~A.,  {Levine} S.,  {Terrell} D.,   {Welch} D.~L.,  2015, in
  American Astronomical Society Meeting Abstracts \#225. p. 336.16

\bibitem[\protect\citeauthoryear{Horvat, Conroy, Pablo, Hambleton, Kochoska,
  Giammarco  \& Pr{\v{s}}a}{Horvat et~al.}{2018}]{Horvat_2018}
Horvat M.,  Conroy K.~E.,  Pablo H.,  Hambleton K.~M.,  Kochoska A.,  Giammarco
  J.,   Pr{\v{s}}a A.,  2018, \mn@doi [The Astrophysical Journal Supplement
  Series] {10.3847/1538-4365/aacd0f}, 237, 26

\bibitem[\protect\citeauthoryear{{Ivezic} \& {Elitzur}}{{Ivezic} \&
  {Elitzur}}{1997}]{DUSTY}
{Ivezic} Z.,  {Elitzur} M.,  1997, \mn@doi [\mnras] {10.1093/mnras/287.4.799},
  \href {https://ui.adsabs.harvard.edu/abs/1997MNRAS.287..799I} {287, 799}

\bibitem[\protect\citeauthoryear{{Jayasinghe} et~al.,}{{Jayasinghe}
  et~al.}{2018}]{tharindu_variable_1}
{Jayasinghe} T.,  et~al., 2018, \mn@doi [\mnras] {10.1093/mnras/sty838}, \href
  {https://ui.adsabs.harvard.edu/abs/2018MNRAS.477.3145J} {477, 3145}

\bibitem[\protect\citeauthoryear{{Jayasinghe} et~al.,}{{Jayasinghe}
  et~al.}{2020}]{2020ATel13745....1J}
{Jayasinghe} T.,  et~al., 2020, The Astronomer's Telegram, \href
  {https://ui.adsabs.harvard.edu/abs/2020ATel13745....1J} {13745, 1}

\bibitem[\protect\citeauthoryear{{Jayasinghe} et~al.,}{{Jayasinghe}
  et~al.}{2021}]{jayasinghe_2021}
{Jayasinghe} T.,  et~al., 2021, \mn@doi [\mnras] {10.1093/mnras/stab114}, \href
  {https://ui.adsabs.harvard.edu/abs/2021MNRAS.503..200J} {503, 200}

\bibitem[\protect\citeauthoryear{{Kochanek} et~al.,}{{Kochanek}
  et~al.}{2017}]{2017PASP..129j4502K}
{Kochanek} C.~S.,  et~al., 2017, \mn@doi [\pasp] {10.1088/1538-3873/aa80d9},
  \href {https://ui.adsabs.harvard.edu/abs/2017PASP..129j4502K} {129, 104502}

\bibitem[\protect\citeauthoryear{{Kov{\'a}cs}, {Zucker}  \&
  {Mazeh}}{{Kov{\'a}cs} et~al.}{2002}]{2002A&A...391..369K}
{Kov{\'a}cs} G.,  {Zucker} S.,   {Mazeh} T.,  2002, \mn@doi [\aap]
  {10.1051/0004-6361:20020802}, \href
  {https://ui.adsabs.harvard.edu/abs/2002A&A...391..369K} {391, 369}

\bibitem[\protect\citeauthoryear{{Kraft}}{{Kraft}}{1967}]{kraft_rapid_rotation}
{Kraft} R.~P.,  1967, \mn@doi [\apj] {10.1086/149359}, \href
  {https://ui.adsabs.harvard.edu/abs/1967ApJ...150..551K} {150, 551}

\bibitem[\protect\citeauthoryear{{Maxted}}{{Maxted}}{2016}]{ellc}
{Maxted} P.~F.~L.,  2016, \mn@doi [\aap] {10.1051/0004-6361/201628579}, \href
  {https://ui.adsabs.harvard.edu/abs/2016A&A...591A.111M} {591, A111}

\bibitem[\protect\citeauthoryear{{Mayor} \& {Mermilliod}}{{Mayor} \&
  {Mermilliod}}{1984}]{mayor_binary_cutoff}
{Mayor} M.,  {Mermilliod} J.~C.,  1984, in {Maeder} A.,  {Renzini} A.,  eds, ~
  Vol. 105, Observational Tests of the Stellar Evolution Theory. p.~411

\bibitem[\protect\citeauthoryear{{Noyes}, {Hartmann}, {Baliunas}, {Duncan}  \&
  {Vaughan}}{{Noyes} et~al.}{1984}]{1984ApJ...279..763N}
{Noyes} R.~W.,  {Hartmann} L.~W.,  {Baliunas} S.~L.,  {Duncan} D.~K.,
  {Vaughan} A.~H.,  1984, \mn@doi [\apj] {10.1086/161945}, \href
  {https://ui.adsabs.harvard.edu/abs/1984ApJ...279..763N} {279, 763}

\bibitem[\protect\citeauthoryear{{Paxton}, {Bildsten}, {Dotter}, {Herwig},
  {Lesaffre}  \& {Timmes}}{{Paxton} et~al.}{2011}]{MESA_1}
{Paxton} B.,  {Bildsten} L.,  {Dotter} A.,  {Herwig} F.,  {Lesaffre} P.,
  {Timmes} F.,  2011, \mn@doi [\apjs] {10.1088/0067-0049/192/1/3}, \href
  {https://ui.adsabs.harvard.edu/abs/2011ApJS..192....3P} {192, 3}

\bibitem[\protect\citeauthoryear{{Paxton} et~al.,}{{Paxton}
  et~al.}{2013}]{MESA_2}
{Paxton} B.,  et~al., 2013, \mn@doi [\apjs] {10.1088/0067-0049/208/1/4}, \href
  {https://ui.adsabs.harvard.edu/abs/2013ApJS..208....4P} {208, 4}

\bibitem[\protect\citeauthoryear{{Paxton} et~al.,}{{Paxton}
  et~al.}{2015}]{MESA_3}
{Paxton} B.,  et~al., 2015, \mn@doi [\apjs] {10.1088/0067-0049/220/1/15}, \href
  {https://ui.adsabs.harvard.edu/abs/2015ApJS..220...15P} {220, 15}

\bibitem[\protect\citeauthoryear{{Pogge} et~al.,}{{Pogge}
  et~al.}{2010}]{PoggeMODS}
{Pogge} R.~W.,  et~al., 2010, in Ground-based and Airborne Instrumentation for
  Astronomy III. p. 77350A, \mn@doi{10.1117/12.857215}

\bibitem[\protect\citeauthoryear{Pr{\v{s}}a \& Zwitter}{Pr{\v{s}}a \&
  Zwitter}{2005}]{Pr_a_2005}
Pr{\v{s}}a A.,  Zwitter T.,  2005, \mn@doi [The Astrophysical Journal]
  {10.1086/430591}, 628, 426

\bibitem[\protect\citeauthoryear{Pr{\v{s}}a et~al.,}{Pr{\v{s}}a
  et~al.}{2016}]{Pr_a_2016}
Pr{\v{s}}a A.,  et~al., 2016, \mn@doi [The Astrophysical Journal Supplement
  Series] {10.3847/1538-4365/227/2/29}, 227, 29

\bibitem[\protect\citeauthoryear{{Ricker} et~al.,}{{Ricker}
  et~al.}{2015}]{tess_instrument_paper}
{Ricker} G.~R.,  et~al., 2015, \mn@doi [Journal of Astronomical Telescopes,
  Instruments, and Systems] {10.1117/1.JATIS.1.1.014003}, \href
  {https://ui.adsabs.harvard.edu/abs/2015JATIS...1a4003R} {1, 014003}

\bibitem[\protect\citeauthoryear{{Scargle}}{{Scargle}}{1982}]{lombscargle}
{Scargle} J.~D.,  1982, \mn@doi [\apj] {10.1086/160554}, \href
  {https://ui.adsabs.harvard.edu/abs/1982ApJ...263..835S} {263, 835}

\bibitem[\protect\citeauthoryear{{Schlafly} \& {Finkbeiner}}{{Schlafly} \&
  {Finkbeiner}}{2011}]{schlafly_finkbeiner_exctinction}
{Schlafly} E.~F.,  {Finkbeiner} D.~P.,  2011, \mn@doi [\apj]
  {10.1088/0004-637X/737/2/103}, \href
  {https://ui.adsabs.harvard.edu/abs/2011ApJ...737..103S} {737, 103}

\bibitem[\protect\citeauthoryear{{Shappee} et~al.,}{{Shappee}
  et~al.}{2014}]{2014ApJ...788...48S}
{Shappee} B.~J.,  et~al., 2014, \mn@doi [\apj] {10.1088/0004-637X/788/1/48},
  \href {https://ui.adsabs.harvard.edu/abs/2014ApJ...788...48S} {788, 48}

\bibitem[\protect\citeauthoryear{{Simonian}, {Pinsonneault}  \&
  {Terndrup}}{{Simonian} et~al.}{2019}]{simonian_rapid_rotation}
{Simonian} G. V.~A.,  {Pinsonneault} M.~H.,   {Terndrup} D.~M.,  2019, \mn@doi
  [\apj] {10.3847/1538-4357/aaf97c}, \href
  {https://ui.adsabs.harvard.edu/abs/2019ApJ...871..174S} {871, 174}

\bibitem[\protect\citeauthoryear{{Strassmeier}, {Hall}, {Zeilik}, {Nelson},
  {Eker}  \& {Fekel}}{{Strassmeier} et~al.}{1988}]{CABs_1}
{Strassmeier} K.~G.,  {Hall} D.~S.,  {Zeilik} M.,  {Nelson} E.,  {Eker} Z.,
  {Fekel} F.~C.,  1988, \aaps, \href
  {https://ui.adsabs.harvard.edu/abs/1988A&AS...72..291S} {72, 291}

\bibitem[\protect\citeauthoryear{{Strassmeier}, {Fekel}, {Bopp}, {Dempsey}  \&
  {Henry}}{{Strassmeier} et~al.}{1990}]{1990ApJS...72..191S}
{Strassmeier} K.~G.,  {Fekel} F.~C.,  {Bopp} B.~W.,  {Dempsey} R.~C.,   {Henry}
  G.~W.,  1990, \mn@doi [\apjs] {10.1086/191414}, \href
  {https://ui.adsabs.harvard.edu/abs/1990ApJS...72..191S} {72, 191}

\bibitem[\protect\citeauthoryear{{Torres} \& {Ribas}}{{Torres} \&
  {Ribas}}{2002}]{mdwarf_problem}
{Torres} G.,  {Ribas} I.,  2002, \mn@doi [\apj] {10.1086/338587}, \href
  {https://ui.adsabs.harvard.edu/abs/2002ApJ...567.1140T} {567, 1140}

\bibitem[\protect\citeauthoryear{{Vallely}, {Kochanek}, {Stanek}, {Fausnaugh}
  \& {Shappee}}{{Vallely} et~al.}{2020}]{Vallely2020}
{Vallely} P.~J.,  {Kochanek} C.~S.,  {Stanek} K.~Z.,  {Fausnaugh} M.,
  {Shappee} B.~J.,  2020, arXiv e-prints, \href
  {https://ui.adsabs.harvard.edu/abs/2020arXiv201006596V} {p. arXiv:2010.06596}

\bibitem[\protect\citeauthoryear{{Watson}, {Henden}  \& {Price}}{{Watson}
  et~al.}{2006}]{vsx_2006}
{Watson} C.~L.,  {Henden} A.~A.,   {Price} A.,  2006, Society for Astronomical
  Sciences Annual Symposium, \href
  {https://ui.adsabs.harvard.edu/abs/2006SASS...25...47W} {25, 47}

\bibitem[\protect\citeauthoryear{{Way} et~al.,}{{Way}
  et~al.}{2019a}]{2019ATel13106....1W}
{Way} Z.,  et~al., 2019a, The Astronomer's Telegram, \href
  {https://ui.adsabs.harvard.edu/abs/2019ATel13106....1W} {13106, 1}

\bibitem[\protect\citeauthoryear{{Way} et~al.,}{{Way}
  et~al.}{2019b}]{2019ATel13159....1W}
{Way} Z.,  et~al., 2019b, The Astronomer's Telegram, \href
  {https://ui.adsabs.harvard.edu/abs/2019ATel13159....1W} {13159, 1}

\bibitem[\protect\citeauthoryear{{Way} et~al.,}{{Way}
  et~al.}{2019c}]{2019ATel13346....1W}
{Way} Z.,  et~al., 2019c, The Astronomer's Telegram, \href
  {https://ui.adsabs.harvard.edu/abs/2019ATel13346....1W} {13346, 1}

\bibitem[\protect\citeauthoryear{{Way} et~al.,}{{Way}
  et~al.}{2021}]{asassn_21co}
{Way} Z.,  et~al., 2021, The Astronomer's Telegram, \href
  {https://ui.adsabs.harvard.edu/abs/2021ATel14436....1W} {14436, 1}

\bibitem[\protect\citeauthoryear{{Zhang}, {Pi}  \& {Zhu}}{{Zhang}
  et~al.}{2015}]{Zhang_chromo}
{Zhang} L.-Y.,  {Pi} Q.-F.,   {Zhu} Z.-Z.,  2015, \mn@doi [Research in
  Astronomy and Astrophysics] {10.1088/1674-4527/15/2/009}, \href
  {https://ui.adsabs.harvard.edu/abs/2015RAA....15..252Z} {15, 252}

\makeatother
\end{thebibliography}








\bsp	
\label{lastpage}
\end{document}